\documentclass{aa}  
\usepackage{graphicx}
\usepackage{xcolor}
\usepackage{upgreek}
\usepackage{mathtools}
\usepackage{txfonts}
\usepackage{hyperref}
\newcommand{\msun}{{\mathrm{M}_\odot}}
\newcommand{\au}{{\mathrm{au}}}

\newcommand{\mbh}{{m}_\mathrm{BH}}

\newcommand{\Rg}{{R}_\mathrm{g}}
\newcommand{\cs}{{c}_\mathrm{s}}
\definecolor{seagreen}{rgb}{0.190, 0.525, 0.361}
\definecolor{cerulean}{rgb}{0.165, 0.322, 0.745}
\definecolor{goldenrod}{rgb}{0.855, 0.647, 0.125}

\newcommand{\REV}[1]{{#1}}

\begin{document}

   \title{Turbulent drag on stellar mass black holes embedded in disks of active galactic nuclei}
   
   \titlerunning{Turbulent drag on stellar mass black holes embedded in AGN disks}
   \authorrunning{A. A. Trani and P. Di Cintio}


   \author{Alessandro Alberto Trani
          \inst{1,2,3}
          \and
          Pierfrancesco Di Cintio\inst{4,5,6}
          }

   \institute{
        Niels Bohr International Academy, Niels Bohr Institute, Blegdamsvej 17, 2100 Copenhagen, Denmark,
        \and
    National Institute for Nuclear Physics – INFN, Sezione di Trieste, I-34127, Trieste, Italy
    \and
    Departamento de Astronom\'ia, Facultad Ciencias F\'isicas y Matem\'aticas, Universidad de Concepci\'on, Concepci\'on, 4030000, Chile
        \email{aatrani@gmail.com}
        \and
        National Council of Research - Institute of Complex Systems, Via Madonna del piano 10, I-50019 Sesto Fiorentino, Italy
        \and
        National Institute of Nuclear Physics (INFN) -  Florence unit, via G. Sansone 1, I-50019 Sesto Fiorentino, Italy
        \and
        National Institute of Astrophysics - Arcetri Astrophysical Observatory (INAF-OAA), Piazzale E.\ Fermi 5, I-50125 Firenze, Italy}

   \date{Received M DD, YYYY; accepted M DD, YYYY}
 
  \abstract
  {Interactions between stellar-mass black holes (BHs) and the accretion disks of supermassive BHs in active galactic nuclei (AGN) constitute a promising channel for the formation of gravitational wave sources. The efficiency of this process depends critically on how embedded BHs evolve under the influence of gaseous drag. Previous studies have assumed laminar disk conditions, leading to idealized configurations with BHs on circular, coplanar orbits. However, AGN disks are expected to be turbulent, and the impact of turbulence on BH orbital evolution remains largely unexplored.}
  {We investigate how AGN disk turbulence affects the orbital dynamics of a stellar-mass BH initially located at a migration trap, focusing on the long-term behavior of eccentricity and inclination in the quasi-embedded regime.}
  {We developed a semi-analytical framework in which turbulence is modeled as a stochastic velocity field acting through a modified drag force. We integrated the resulting stochastic differential equations both in Cartesian coordinates and in orbital elements using a linearized perturbative approach and compared these results with full numerical simulations.}
  {Eccentricity and inclination evolve toward steady-state Rayleigh distributions, with variances determined by the local disk properties and the ratio of the gas damping rate to the orbital frequency. The analytical predictions agree well with the numerical simulations. We provide closed-form expressions for the variances in both the fast and slow damping regimes. These results are directly applicable to Monte Carlo population models and can serve as physically motivated initial conditions for hydrodynamical simulations.}
  {Turbulent forcing prevents full circularization and alignment of BH orbits in AGN disks, even in the presence of strong gas drag. This has important implications for BH merger and binary formation rates, which are sensitive to the residual eccentricity and inclination. Our results highlight the need to account for turbulence-induced stochastic heating when modeling the dynamical evolution of compact objects in AGN environments.}
   \maketitle
%
\section{Introduction}

Gravitational waves (GWs) are rapidly becoming a powerful tool to probe the complex dynamics of dense stellar environments and the physics of compact objects lurking at their centers \citep{mapelli2021review,speratrani2022,gwtc-3pop,KAGRA:2021vkt}. Upcoming ground-based detectors such as the Einstein Telescope \citep[ET,][]{punturo2010} and the Cosmic Explorer \citep[CE,][]{reitze2019} will significantly extend sensitivity to lower frequencies and larger volumes, enabling the detection of black hole (BH) mergers out to high redshift and providing improved access to the early inspiral phase of stellar-mass binaries \citep{ET:2019dnz,branchesi2023}. Complementing these efforts, future space-borne GW detectors such as LISA \citep{2017arXiv170200786A} or TAIJI \citep{10.1093/nsr/nwx116} will open the sub-Hz band, allowing the detection of long-lived inspirals involving both supermassive BH (SMBH) binaries and stellar-mass BHs in extreme mass-ratio configurations \citep{hopman2005,pau2007,mandel2008,babak2017}. 

As GW detectors reach higher sensitivity, attention is turning to the role of environmental effects in shaping binary evolution \citep{zwick2024,zwick2025,tatatsky2025}. Among these, gas-rich environments such as active galactic nuclei (AGNs) offer particularly promising prospects. Interactions with AGN accretion disks can significantly alter the dynamics of embedded BH binaries through hydrodynamic drag, accretion torques, and disk-driven migration, potentially accelerating inspiral or exciting orbital eccentricity \citep{stone2016,ishibashi2020,li2021,li2022,dittmann2022,samsing2022,li2023a,li2023b,rowan2023,rowan2024a,rowan2024b,whitehead2024,whitehead2024b,li2024a,dittmann2024,dodici2024,bhupendra2024,trani2024a,whitehead2025,whitehead2025b,dittmann2025,rowan2025}. These effects may leave detectable imprints on the GW signal, opening a window onto the astrophysical environments in which BH mergers occur. Other mechanisms, such as dynamical friction from dark matter spikes, have also been thought to influence binary evolution in galactic nuclei  \citep[e.g.,][]{2020PhRvD.102j3022H,2024Univ...10..427M,2024MNRAS.533.2335M,2024PhRvD.110h3027D,2024A&A...690A.299F,2024arXiv240213762K}. However, AGN-assisted mergers are particularly compelling because they may also produce electromagnetic counterparts through shocks, variable accretion, or relativistic jets, thus enabling multi-messenger observations \citep{graham2020,ren2022,gayathri2023,tagawa2023,chen2024,tagawa2024}.

Magnetohydrodynamical simulations have revealed that disk dynamics, ranging from protoplanetary to AGN scales, are largely governed by turbulence driven by magnetorotational instability \citep{balbus1998,2004ApJ...602..595J,armitage2011,2024A&A...691A.151Z}. This turbulence can induce substantial density and velocity fluctuations, which can have a strong impact on the orbital evolution of embedded objects \citep{papaloizou2004,nelson2004,nelson2005,johnson2006,2007ApJ...670..805O,yang2009,nelson2010,yang2012}.

From the perspective of kinetic (i.e., particle based) simulations, only limited attention has been paid to the impact of the density fluctuations on the orbit of objects embedded in turbulent gaseous disks \citep{rein2009,baruteau2010,secunda2019}, and even less effort has been devoted to modeling the direct impact of turbulent velocity fields. A key obstacle lies in the broad range of scales involved: hydrodynamic turbulence evolves rapidly on small spatial scales, whereas orbital motion proceeds more slowly and over larger distances. This disparity makes it computationally challenging to simultaneously resolve both the trajectory of a compact object and the collective dynamics of the surrounding disk within a single hybrid simulation. 

In this work, we investigated the role of turbulence in shaping the orbital evolution of a stellar-mass BH embedded in the accretion disk of an SMBH, focusing on orbits near the migration trap radius. To this end, we developed a semi-analytical framework based on a stochastic differential equation that captures turbulent velocity fluctuations through a time-varying drag coefficient.

The paper is organized as follows. In Sect.~\ref{models} we introduce the AGN disk model, the governing equations, and the numerical techniques employed. In Sect.~\ref{results} we present the results of our semi-analytical calculations and construct a reduced model to interpret the key features. Finally, in Sect.~\ref{discussion} we summarize our findings and discuss their astrophysical implications.
\begin{figure}
        \centering
        \includegraphics[width=0.98\linewidth]{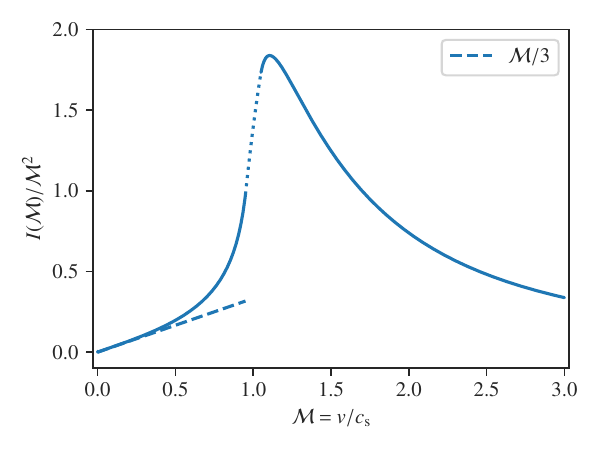}
        \caption{Dependency of \citet{ostriker1999} the dynamical friction prescription as a function of the mach number $\mathcal{M}$ (Eq.~\ref{eq:prescription}). The dotted line marks the linear interpolation used to circumvent the mathematical discontinuity at $\mathcal{M} = 1$. The dashed line indicates the linear subsonic limit. 
        }
        \label{fig:idrag}
\end{figure}

\section{Models}\label{models}

\subsection{The turbulent AGN disk model}

In our kinetic model, we account for the effect of a turbulent AGN disk via an effective friction and diffusion process. The equation of motion for the secondary BH of stellar mass ($m_{\rm BH}$) orbiting in the gravitational field of the SMBH $M_{\rm BH}$ reads
\begin{equation}\label{eq:eomcart}
\frac{{\rm d^2}\mathbf{r}}{{\rm d}t^2}=-\frac{GM_{\rm BH}}{r^3}\mathbf{r}-\eta(\tilde{v},R,z)\tilde{\mathbf{v}},
\end{equation}
where $\eta$ is the local (i.e., dependent on the cylindrical radial position and vertical distance $z$) friction coefficient, accounting for the drag force exerted by the disk. The relative velocity, $\tilde{\mathbf{v}}$, is defined as  
\begin{equation}
\tilde{\mathbf{v}}=\mathbf{v}-\mathbf{v}_{\rm circ}-\mathbf{v}_{\rm turb}. 
\end{equation}
In the equation above, $\mathbf{v}=\dot{\mathbf{r}}$; $\mathbf{v}_{\rm circ}$ is the gas circular velocity at cylindrical radius $R$, of magnitude $v_{\rm circ}=\sqrt{GM_{\rm BH}/R}$. $\mathbf{v}_{\rm turb}$ is the stochastic fluctuation of $\mathbf{v}_{\rm circ}$ induced by the turbulent gas motion in the disk. The drag coefficient is defined by 
\begin{equation}\label{eq:eta}
\eta=4\pi G^2\rho m_{\rm BH} \frac{I(\mathcal{M})}{\tilde{v}^3},    
\end{equation}
where the quantity $I$ is given as a function of the local Mach number $\mathcal{M}=\tilde{v}/\cs$ by the \cite{ostriker1999} prescription
\begin{equation}\label{eq:prescription}
I=
\begin{cases}
\displaystyle \frac{1}{2}\log\frac{1+\mathcal{M}}{1-\mathcal{M}}-\mathcal{M}\quad \mathcal{M}\leq 0.95,\\
\displaystyle \frac{1}{2}\log\Big( 1-\frac{1}{\mathcal{M}^2}\Big)+3.1\quad \mathcal{M}\geq 1.05,\\
\displaystyle 0.88+10.3\Big(\mathcal{M}-0.95\Big)\quad {\rm elsewhere}. 
\end{cases}
\end{equation}
In the definition above, the third expression is a linear interpolation between the subsonic and supersonic regimes\footnote{Note that, in the limit of $\mathcal{M}\gg 1$, one recovers the classical expression for the \cite{chandrasekhar1943} dynamical friction in a collisionless gravitational system of mean mass density $\rho$.} to overcome the mathematical discontinuity at $\mathcal{M}=1$ (see Fig.~\ref{fig:idrag} and \citealt{delaurentis2023}).

We note that the model given by Eq.~(\ref{eq:eomcart}) cannot be reduced in terms of a simple second order Langevin equation (see, e.g., the discussion in \citealt{2025arXiv250322479S}) of the form  
\begin{equation}\label{eq:langeq}
\ddot{\mathbf{r}}=-\nabla\Phi-\eta\dot{\mathbf{r}}+\delta \mathbf{f},
\end{equation}
where $\delta \mathbf{f} $ is a fluctuating force per unit mass, as the velocity fluctuations induced by turbulence also enter explicitly the definition of the friction coefficient $\eta$.

\begin{figure}
        \centering
        \includegraphics[width=1\linewidth]{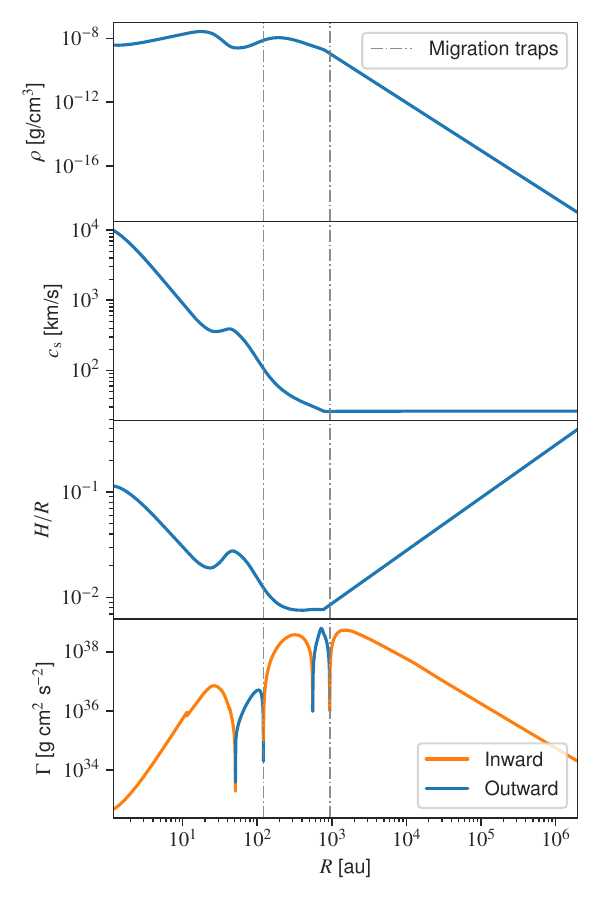}
        \caption{Radial profiles of the AGN disk employed in this work. From top to bottom: \REV{midplane} gas density $\rho$, sound speed $\cs$, disk aspect ratio $h/R$, and magnitude of the migration torque $\Gamma$. The migration torque assumes a secondary BH mass of $\mbh = 20 \,\msun$. The outer migration trap lies within the star formation region, which begins at $R \simeq 767 \,\au$.}
        \label{fig:sirko1e7}
\end{figure}

In this work we neglected the gravitational field of the AGN disk and assumed an axisymmetric density $\rho(R,z)$ of the form
\begin{equation}\label{eq:rhodisk}
\rho(R,z)=\rho_R\exp\left(-\frac{z^2}{2H_R^2}\right),
\end{equation}
where $\rho_R$ and $H_R$ are the \REV{midplane} density and scale height at cylindrical radius $R$, respectively. In the numerical simulations discussed hereafter, we interpolated its parameters from a 1D AGN model generated with \texttt{pagn} \citep{gangardt2024}. We adopted the \citet{sirko2003} model with $M_\mathrm{BH} = 10^7 \, \msun$ and the alpha viscosity $\alpha = 0.01$, the Eddington luminosity ratio $l_{E} = 0.5$, and the radiative efficiency $\varepsilon = 0.1$. The density profile of the disk was truncated at $R=2\times10^7\, \Rg$. Throughout this work we assumed $m_{\rm BH}=10\, \msun$. Figure~\ref{fig:sirko1e7} shows the radial density \REV{in the midplane of the AGN disk (top panel, corresponding to $\rho_R$ in Eq.~\ref{eq:rhodisk})}, its sound speed and scale height profiles (mid-panels), as well as the magnitude of the so-called migration torque $\Gamma$ (bottom panel), defined as the effective force exerted on $m_{\rm BH}$ by the leading and trailing spiral density waves induced by the perturbations of the disk density. The migration torque is evaluated as in \citet[][see also \citealt{masset2017,guilera2021,grishin2024}]{gangardt2024}. The model disk has two migration traps, $R_{{\rm trap},1}\simeq 121 \,\au \simeq 1230\,\Rg$ and $R_{{\rm trap},2}\simeq 939 \,\au \simeq 9512\,\Rg$, indicated in the figure by the vertical dashed lines. They are defined as the radii at which the torque $\Gamma$ is such that, for $R<R_{{\rm trap},i}$, the particle $m_{\rm BH}$ is pushed outward, and inward otherwise. For reasons of simplicity, as we only considered initial conditions starting at migration traps, we neglected the effect of the migration torque in our model.

\subsection{Governing equations}
Starting from Eq.~(\ref{eq:eomcart}) in Cartesian coordinates, we derived the Gaussian perturbation equations in terms of Keplerian orbital elements using standard perturbative techniques \citep[see][]{beutler2005}. All numerical results in this work are based on the full form of these equations, which we independently validated by comparing with direct integrations of Eq.~(\ref{eq:eta}) in Cartesian coordinates (see Sect.~\ref{sec:numerics}). To develop analytical insight, however, we also considered simplified versions of the perturbation equations that capture the essential physics while remaining more tractable. Refer to Table~\ref{tab:symbols} for the full list of symbols used in this manuscript.

Rather than using the full expression for $\eta(\tilde{v})$, which depends on a complex nonlinear dependence on the relative velocity, we took advantage of the fact that the orbits under consideration are fully embedded within the AGN disk. These orbits typically have low eccentricities ($e$) and inclinations ($\iota$). To first order in $e$ and $\iota$, the relative velocity (excluding contributions from the turbulence) are approximated as
\begin{equation}\label{eq:veldiff}
    \tilde{v}^2 \simeq v_\mathrm{circ}^2 \left[ e^2 \left(1 - \frac{3}{4} \cos^2{\upnu}\right) + \iota^2 \cos^2{u} \right],
\end{equation}
where $\upnu$ is the true anomaly and $u = \omega + \upnu$ is the argument of latitude. For a stellar-mass BH embedded in the disk with $e, \iota \ll 1$, the relative velocity satisfies $\tilde{v} \ll v_\mathrm{circ} = c_\mathrm{s} (H / R)^{-1}$, where the expression for $v_\mathrm{circ}$ follows from vertical hydrostatic equilibrium. 

Given that the disk aspect ratio satisfies $H / R \gtrsim 0.01$ everywhere (see Fig.~\ref{fig:sirko1e7}), we obtain $\tilde{v} \ll \cs$, and the gas friction is always subsonic. In this regime, for Mach numbers $\mathcal{M} \ll 1$ (a condition that holds up to $\mathcal{M} \simeq 0.5$; see Fig.~\ref{fig:idrag}), the drag coefficient in Eq.~(\ref{eq:eta}) becomes independent of $\tilde{v}$ and simplifies to
\begin{equation}\label{eq:etastar}
\eta_* = \frac{4\pi G^2 \rho \,\mbh}{3 \cs^3}.
\end{equation}
Here, $\rho$ retains its dependence on the radial and vertical coordinates $R$ and $z$, as given by Eq.~(\ref{eq:rhodisk}). For convenience, we also define the associated damping timescale $\tau_*$:
\begin{equation}\label{eq:taustar}
        \tau_* = \frac{1}{\eta_*} = \frac{3 \cs^3}{4\pi G^2 \rho \,\mbh}.
\end{equation}

Consequently, in the subsonic regime, the drag force becomes linear in the velocity $\tilde{v}$, allowing us to decompose the acceleration appearing in Eq. (\ref{eq:eomcart}) into the deterministic component
\begin{equation}\label{eq:adet}
\left.\frac{{\rm d^2}\mathbf{r}}{{\rm d}t^2}\right|_{\rm det} = -\frac{GM_{\rm BH}}{r^3}\mathbf{r}-\eta_* (\mathbf{v}-\mathbf{v}_\mathrm{circ}),    
\end{equation}
and the stochastic fluctuating term
\begin{equation}\label{eq:asto}
\left.\frac{{\rm d^2}\mathbf{r}}{{\rm d}t^2}\right|_{\rm sto} = \eta_* \mathbf{v}_{\rm turb}.    
\end{equation}
In practice, one can collect $\eta_*$ and $\mathbf{v}_{\rm turb}$ in Eq. (\ref{eq:asto}) into an effective fluctuating force $\delta\mathbf{f}_*$, thereby basically recovering Eq. (\ref{eq:langeq}).

Using standard first-order perturbation theory, the Gaussian equations governing the evolution of the semimajor axis ($a$), the eccentricity ($e$), the inclination ($\iota$), the argument of pericenter ($\omega$), the longitude of pericenter ($\varpi = \omega + \Omega$), and the longitude of ascending node ($\Omega$) derived from the deterministic component of the equations Eq.~(\ref{eq:adet}) read
\begin{align}\label{eq:fullpert}
\frac{{\rm d}a}{{\rm d}t}= & -\frac{2 a \eta_*}{1-e^2}  \left(1 + e^2 + 2e\cos\upnu - \frac{\cos\iota \, (1 + e \cos\upnu)^{3/2}}{(\cos^2\iota + \cos^2{u} \sin^2\iota)^{3/4}}\right),\\
\frac{{\rm d}e}{{\rm d}t} = & -\eta_* \left(\cos\upnu + \frac{e+\cos\upnu}{1+e \cos \upnu} \right)\cdot \notag\\
&  \left(1+e\cos\upnu-\frac{\cos\iota \,(1 + e \cos\upnu)^{1/2} }{(\cos^2\iota + \cos^2{u} \sin^2\iota)^{3/4}}\right) - e \eta_*\sin^2\upnu, \\
\frac{{\rm d}\iota}{{\rm d}t} = & -\eta_*\frac{\cos^2{u}\,\sin\iota}{(1+e \cos\upnu)^{1/2} (\cos^2{u} + \cos^2{\iota} \sin^2{u})^{3/4}}, \\
\frac{{\rm d}\varpi}{{\rm d}t} = & -\eta_* \frac{\sin{u}}{e}\left(2 - \frac{\cos\iota\,(2 + e\cos\upnu)}{(1 + e \cos\upnu)^{1/2} (\cos^2\iota + \cos^2{u} \sin^2\iota)^{3/4}}\right), \\
\frac{{\rm d}\Omega}{{\rm d}t} = & -\eta_* \frac{\sin{u}\cos{u}}{(1+e\cos\upnu)^{1/2}(\cos^2{u} + \cos^2{\iota} \sin^2{u})^{3/4}}.\label{eq:lastpert}
\end{align}
This system is closed by the usual two-body expressions for ${\rm d}\upnu/{\rm d}t$ and ${\rm d}\omega/{\rm d}t$, while ensuring that the reference frame contributions are taken into account \citep[e.g.,][]{burns1976}:
\begin{align}
        \frac{{\rm d}\upnu}{{\rm d}t} = & \sqrt{\frac{\mu}{a^3}} \frac{(1 + e \cos{\upnu})^2}{(1-e^2)^{3/2}} - \frac{{\rm d}\varpi}{{\rm d}t},\\
        \frac{{\rm d}\omega}{{\rm d}t} = & \frac{{\rm d}\varpi}{{\rm d}t} -\cos{\iota}\frac{{\rm d}\Omega}{{\rm d}t}.
\end{align}
Here, $\mu = GM_{\rm BH}$ denotes the standard gravitational parameter.

To gain analytical insight, we expanded the perturbation equations to first order in $e$ and $\iota$. The resulting linearized system for the orbital elements reads:
\begin{align}\label{eq:linear}
& \frac{{\rm d}a}{{\rm d}t}= - \eta_* a e \cos{u}, \\
& \frac{{\rm d}e}{{\rm d}t}= - \eta_* e, \label{eq:epert} \\
& \frac{{\rm d}\iota}{{\rm d}t}= -\eta_* \iota \cos^2{u}, \label{eq:iotapert} \\
& \frac{{\rm d}\varpi}{{\rm d}t} = \frac{1}{4} \eta_* e \cos^2{\upnu} \sin{\upnu}, \\
& \frac{{\rm d}\Omega}{{\rm d}t} = \frac{1}{2} \eta_* (\cos{\upnu} - 2) \sin{u}\cos{u}. \label{eq:lastlinear}
\end{align}

To model the stochastic component of the orbital evolution, we assumed that the turbulent velocity perturbation, $\mathbf{v}_{\rm turb}$, follows a stationary stochastic Gaussian process with an autocorrelation function given by
\begin{equation}
        \langle v_{\rm turb}(t_1)\,v_{\rm turb}(t_2) \rangle = G(t_1-t_2) = \sigma_{\rm turb}^2\, \exp{\left(-\frac{|t_1 - t_2|}{\tau_\mathrm{c}}\right)},
\end{equation}
where $\sigma_{\rm turb}^2 \equiv \langle v^2_\mathrm{turb} \rangle$ is the variance of the turbulent velocity field, and $\tau_\mathrm{c} = 1/\Omega_{\rm circ}$ is the autocorrelation time, with $\Omega_{\rm circ}$ denoting the local gas circular frequency at the position of the stellar-mass BH $m_{\rm BH}$. This assumption is consistent with previous models of turbulence in protoplanetary disks \citep[e.g.,][]{rein2009, picogna2018}.
The amplitude of $\mathbf{v}_{\rm turb}$ is directly related to the local kinematic viscosity of the disk, which within the framework of $\alpha$-disks, is given by
\begin{equation}\label{diffusion}
\nu_{\rm kin} = \alpha \cs H.
\end{equation}
It follows that the 1D variance of the turbulent velocity field $\sigma_\mathrm{turb}^2$ field reads
\begin{equation}\label{eq:sigmaturb}
\sigma_\mathrm{turb}^2 = \frac{\nu_{\rm kin}}{\tau_\mathrm{c}} = \alpha \cs H \,\Omega_{\rm circ} = \alpha \cs^2,
\end{equation}
where the last identity uses $H = \cs / \Omega_{\rm circ}$, which follows from vertical hydrostatic equilibrium. 

The turbulent velocity vector in Cartesian coordinates is expressed as
  \begin{align}\label{eq:vturb}
        \mathbf{v}_{\rm turb} &= \begin{pmatrix}
                v_{{\rm turb}, r} \cos{\phi} - v_{{\rm turb}, \phi} \sin{\phi} \\
                v_{{\rm turb}, r} \sin{\phi} + v_{{\rm turb}, \phi} \cos{\phi} \\
                v_{{\rm turb}, z}
        \end{pmatrix},
\end{align}
where $\phi$ is the cylindrical azimuth of the BH, i.e., the angle in the disk plane from the $+x$ axis to the projection of the BH position onto the $x$--$y$ plane, such that $v_{{\rm turb}, r}$ and $v_{{\rm turb}, \phi}$ are the radial and azimuthal components of the turbulent velocity at the BH location, and $v_{{\rm turb}, z}$ is the vertical component. We assumed that the radial, azimuthal, and vertical components of $\mathbf{v}_{\rm turb}$ were independent, stationary Gaussian processes, each with variance $\sigma_\mathrm{turb}^2$.

\begin{table}[tb]
        \centering
        \caption{Symbols used in the manuscript.}
        \begin{tabular}{ll}
                \hline
                \textbf{Symbol} & \textbf{Description} \\
                \hline
                $a$ & Semimajor axis \\
                $e$ & Eccentricity \\
                $\iota$ & Inclination \\
                $\omega$ & Argument of pericenter \\
                $\Omega$ & Longitude of ascending node \\
                $\varpi$ & Longitude of pericenter ($\omega + \Omega$) \\
                $\upnu$ & True anomaly \\
                $u$ & Argument of latitude ($\omega + \upnu$) \\
                $\mu$ & Gravitational parameter ($G M_{\rm BH}$) \\
                $\cs$ & Sound speed \\
                $\alpha$ & Disk viscosity parameter \\
                $H$ & Disk scale height \\
                $R$ & Cylindrical radial coordinate \\
                $z$ & Vertical coordinate \\
                $\eta$ & General drag coefficient \\
                $\eta_*$ & Subsonic-limit drag coefficient \\
                $\mathbf{v}$ & Velocity of the object \\
                $\mathbf{v}_{\rm circ}$ & Circular velocity of the disk gas \\
                $\mathbf{v}_{\rm turb}$ & Turbulent velocity perturbation \\
                $\sigma_{\rm turb}$ & Turbulent velocity dispersion \\
                $\tau_{\rm c}$ & Autocorrelation time of turbulence \\
                \hline
        \end{tabular}
        \label{tab:symbols}
\end{table}

Deriving the Gaussian perturbation equations corresponding to the stochastic acceleration term in Eq.~(\ref{eq:asto}) is cumbersome but straightforward. The full set of equations reads:
\begin{align}\label{eq:fullturb}
        \frac{{\rm d}a}{{\rm d}t} = & \eta_* \sqrt{\frac{ a^3}{\mu(1-e^2)}} \Bigg\{ \left[ e \sin\upnu \,(\cos^2\iota + \cos^2{u} \sin^2\iota)^{1/2} \right] v_{{\rm turb}, r} + \notag\\
    & + \frac{\cos\iota \, (1 + e\cos\upnu)}{(\cos^2{u} + \cos^2\iota \sin^2{u})^{1/2}} v_{{\rm turb}, \phi} + \notag\\
    & + \sin\iota \,(e\cos\omega + \cos{u}) \,v_{{\rm turb}, z} \Bigg\}, \\
        \frac{{\rm d}e}{{\rm d}t}=  &\eta_* \sqrt{\frac{a (1-e^2)}{\mu}} \Bigg\{ \Bigg[ \sin\upnu \,(\cos^2\iota + \cos^2{u} \sin^2\iota)^{1/2} + \notag\\& - \left(\cos\upnu + \frac{e+\cos\upnu}{1+e\cos\upnu}\right) \frac{\sin^2\iota\sin{u}\cos{u}}{(\cos^2{u} + \cos^2\iota \sin^2{u})^{1/2}} \Bigg] v_{{\rm turb}, r} + \notag\\
        & + \left[\frac{\cos\iota}{(\cos^2{u} + \cos^2\iota \sin^2{u})^{1/2}} \, \left(\cos\upnu +  \frac{e+\cos\upnu}{1+e\cos\upnu}\right)\right] v_{{\rm turb}, \phi} + \notag\\
        & + \Bigg[\sin\upnu\sin{u} +  \left(\cos\upnu+\frac{e+\cos\upnu}{1+e\cos\upnu}\right)\cos{u}\Bigg] \sin\iota\, v_{{\rm turb}, z} \Bigg\}, \\
        \frac{{\rm d}\iota}{{\rm d}t} = & -\eta_* \sqrt{\frac{a (1-e^2)}{\mu}} \frac{\cos{u}}{(1+e\cos\upnu)(\cos^2{u} + \cos^2\iota \sin^2{u})^{1/2}} \cdot \notag\\ &\Bigg\{\sin\iota\cos\iota \sin{u}\,v_{{\rm turb}, r} + \sin\iota\cos{u} \,v_{{\rm turb}, \phi} + \notag\\ 
        & - (\cos^2{u} + \cos^2\iota \sin^2{u})^{1/2}\cos\iota \, v_{{\rm turb}, z} \Bigg\}, \\ 
        \frac{{\rm d}\varpi}{{\rm d}t} = & \frac{\eta_*}{e} \sqrt{\frac{a (1-e^2)}{\mu}} \Bigg\{-\Bigg[\cos\upnu (\cos^2\iota + \cos^2{u} \sin^2\iota)^{1/2} + \\ &+ \left(\frac{2+e\cos\upnu}{1+e\cos\upnu}\right)\frac{\sin^2\iota\sin\upnu\sin{u}\cos{u}}{(\cos^2{u} + \cos^2\iota \sin^2{u})^{1/2}}\Bigg]v_{{\rm turb}, r} + \notag\\
        & + \left[\left(\frac{2+e\cos\upnu}{1+e\cos\upnu}\right)\frac{\cos\iota\sin\upnu}{(\cos^2{u} + \cos^2\iota \sin^2{u})^{1/2}}\right] v_{{\rm turb}, \phi} + \notag\\ 
        & \left[\left(\frac{2+e\cos\upnu}{1+e\cos\upnu}\right)\cos{u}\sin\upnu - \cos\upnu\sin{u}\right] \sin\iota \,v_{{\rm turb}, z} \Bigg\}, \\ 
        \frac{{\rm d}\Omega}{{\rm d}t} = & -\eta_* \sqrt{\frac{a (1-e^2)}{\mu}} \frac{\sin{u}}{1+e\cos\upnu}  \Bigg[ \frac{\cos\iota\sin{u} \,v_{{\rm turb}, r} }{(\cos^2{u} + \cos^2\iota \sin^2{u})^{1/2}}  + \notag\\  & + \frac{\cos{u}}{(\cos^2{u} + \cos^2\iota \sin^2{u})^{1/2}}v_{{\rm turb}, \phi} - \frac{\cos\iota}{\sin\iota}v_{{\rm turb}, z} \Bigg]. \label{eq:lastturb} 
\end{align}

We could, in principle, linearize Eqs.~(\ref{eq:fullturb}--\ref{eq:lastturb}) in $e$ and $\iota$, following the same procedure used for the deterministic system in Eqs.~(\ref{eq:fullpert}--\ref{eq:lastpert}). However, the turbulent velocity dispersion $\sigma_\mathrm{turb}$ is itself of order $\mathcal{O}(e, \iota)$, as shown in Eqs.~(\ref{eq:veldiff}) and (\ref{eq:sigmaturb}). Therefore, to ensure consistency with the perturbative order of the deterministic equations, we expand Eqs.~(\ref{eq:fullturb}--\ref{eq:lastturb}) to zeroth order in eccentricity and inclination:
\begin{align}\label{eq:linearturb}
& \frac{{\rm d}a}{{\rm d}t}= 2a \, \eta_* \sqrt{\frac{a}{\mu}} v_{{\rm turb}, \phi},\\
& \frac{{\rm d}e}{{\rm d}t}= \eta_*  \sqrt{\frac{a}{\mu}} \left(v_{{\rm turb}, r} \sin\upnu + 2 v_{{\rm turb}, \phi} \cos\upnu \right), \label{eq:eturb}\\
& \frac{{\rm d}\iota}{{\rm d}t}= \eta_* \sqrt{\frac{a}{\mu}} \cos{u} \, v_{{\rm turb}, z}, \label{eq:iotaturb}\\
& \frac{{\rm d}\varpi}{{\rm d}t} = \frac{\eta_*}{e} \sqrt{\frac{a}{\mu}} \left(- \cos\upnu \,v_{{\rm turb}, r} + 2 \sin\upnu \, v_{{\rm turb}, \phi} \right), \\
& \frac{{\rm d}\Omega}{{\rm d}t} =  \frac{\eta_*}{i} \sqrt{\frac{a}{\mu}} \sin{u} \,v_{{\rm turb}, z}. \label{eq:lastlinearturb} 
\end{align}
A numerical integration of the reduced stochastic system given by Eqs.~(\ref{eq:linearturb}--\ref{eq:lastlinearturb}) alongside the linearized deterministic system in Eqs.~(\ref{eq:linear}--\ref{eq:lastlinear}) yields results that are virtually indistinguishable from those obtained using the full equations, Eqs.~(\ref{eq:fullpert}--\ref{eq:lastpert}) and (\ref{eq:fullturb}--\ref{eq:lastturb}). Furthermore, the latter reduced equations remain valid across the entire Mach number range, provided that the drag coefficient $\eta_*$ in Eq.~(\ref{eq:etastar}) is replaced with its general expression $\eta$ from Eq.~(\ref{eq:eta}).

\begin{figure*}
        \centering
        \includegraphics[width=\linewidth]{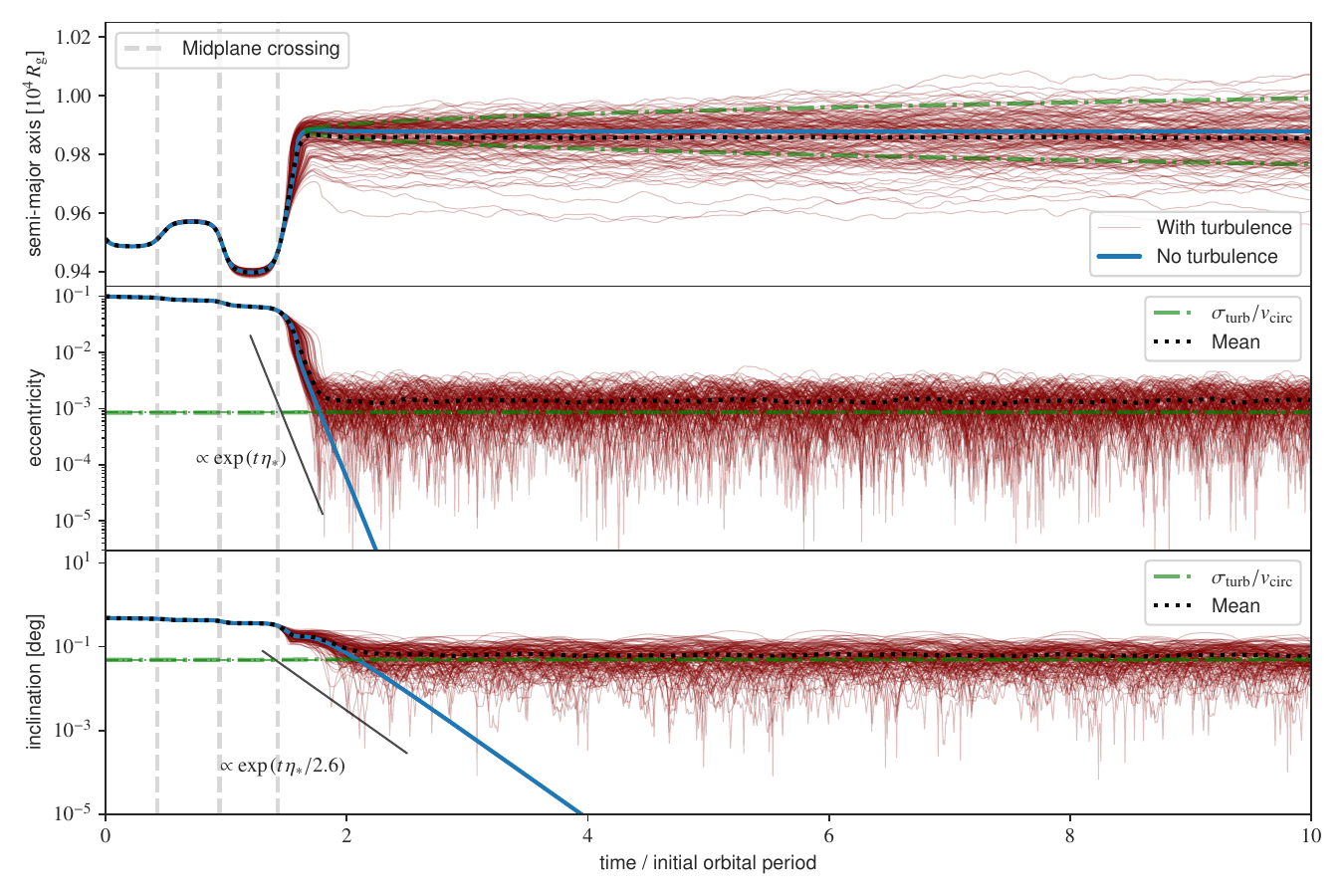}
        \caption{Evolution of the semimajor axis (top), the eccentricity (middle), and the inclination (bottom) for an inclined ($\iota_0 = 0.5^\circ$) and eccentric ($e_0 = 0.1$) BH ($\mbh = 20 \,\msun$) undergoing dynamical friction in the AGN disk. The thick blue line indicates the evolution without the turbulent velocity field, while each thin red line includes a different realization of the turbulent velocity field. The BH is placed at the migration trap within an AGN disk around a $10^7 \, \msun$ SMBH. The dotted black line is the mean evolution of the realizations including turbulence. In the top panel, the dot-dashed lines indicate the mean square change in semimajor axis $\langle \Delta a^2 \rangle$ (Eq.~\ref{eq:deltaa}). In the middle and bottom panels, the dot-dashed green line is the ratio between the turbulent velocity dispersion $\sigma_\mathrm{turb}$ and the circular velocity $v_\mathrm{circ}$. The vertical dashed lines indicate the first three midplane crossings. The turbulence prevents the full circularization and alignment of the embedded BH.
        }
        \label{fig:evolturb}
\end{figure*}
\begin{figure}
        \centering
        \includegraphics[width=\columnwidth]{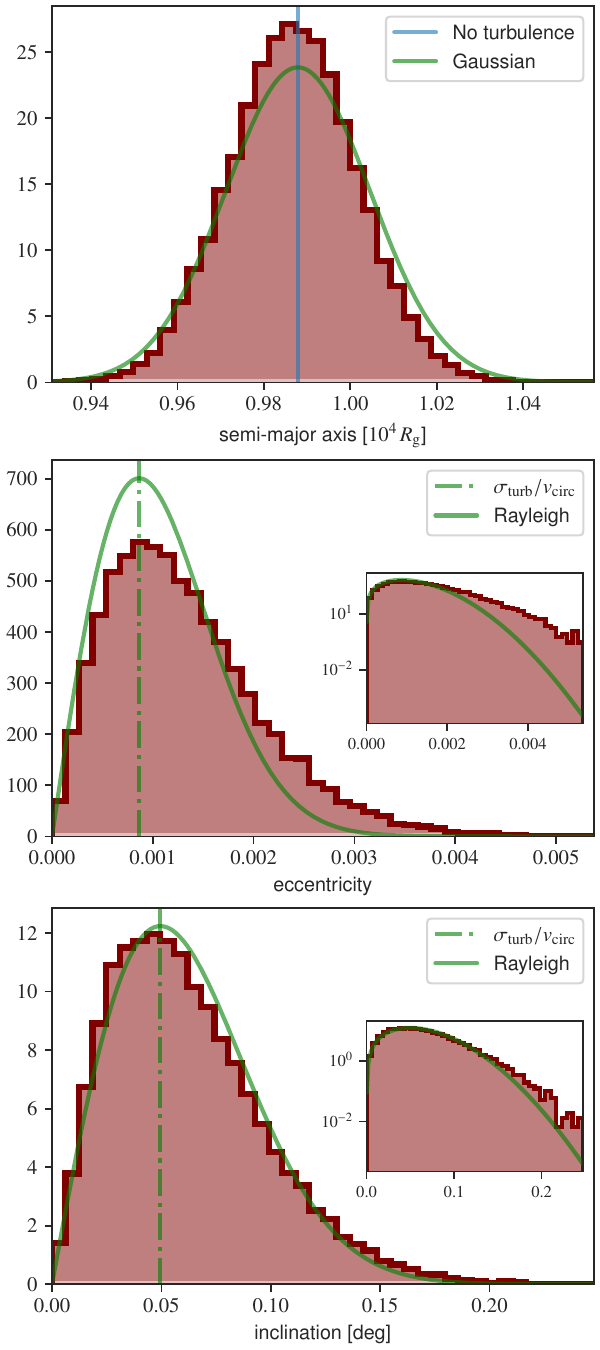}
        \caption{Distributions (normalized to one) of the semimajor axis (top), the eccentricity (middle) and the inclination (bottom) for $2.5\times 10^4$ realizations of an embedded BH, evaluated after 20 initial orbital periods. The initial conditions are identical to those in Fig.~\ref{fig:evolturb}. In the top panel, the blue line marks the final semimajor axis from the nonturbulent simulation, while the green curve shows a Gaussian distribution with standard deviation given by Eq.~(\ref{eq:deltaa}). In the middle and bottom panels, the dot-dashed green line is the ratio between the turbulent velocity dispersion $\sigma_\mathrm{turb}$ and the circular velocity $v_\mathrm{circ}$, while the green curves indicate Rayleigh distributions with mean value $\sigma_\mathrm{turb} / v_\mathrm{circ}$. The insets show the same distributions in logarithm scale, highlighting the presence of fatter tails with respect to the Rayleigh prediction. 
        }
        \label{fig:dist}
\end{figure}

\subsection{Numerical methods}\label{sec:numerics}

In the numerical simulations discussed in the next section, we integrated the system of first-order differential equations~(Eqs. \ref{eq:fullpert}--\ref{eq:lastpert}) governing the evolution of the Keplerian elements using an eighth-order Dormand–Prince scheme \citep{hairer1993}. The drag coefficient $\eta_*$ was replaced with its general form $\eta(\tilde{v}, R, z)$ to account for full velocity and spatial dependence. We employed an adaptive timestep, $\delta t$, with a relative tolerance of $10^{-8}$.
Turbulent velocity fluctuations were modeled as a 3D Gaussian random process with isotropic velocity dispersion $\sigma_{\rm turb} = \sqrt{\alpha}\, \cs$. The temporal correlation of the fluctuations is characterized by the correlation time $\tau_\mathrm{c}$, as introduced above (see \citealt{2003astro.ph.12434T,2004ApJ...602..678S} for further discussion). To implement this, we defined a decay factor $f_{\delta t} = \exp(-\delta t / \tau_\mathrm{c})$, such that at each timestep $k+1$, each component of the turbulent velocity vector $\mathbf{v}_{\rm turb}$ was independently updated from its previous value at step $k$ according to
\begin{equation}\label{eq:vturbiter}
v_{\rm turb}^{k+1} = v_{\rm turb}^k f_{\delta t}+\sqrt{1-f_{\delta t}^2}\, \mathcal{G}(\sigma_{\rm turb}),
\end{equation}
where $\mathcal{G}(\sigma_{\rm turb})$ denotes a random variate drawn from a Gaussian distribution with zero mean and standard deviation $\sigma_{\rm turb}$, generated at each timestep using the Box–Muller algorithm \citep[see, e.g.,][]{2002nrca.book.....P}.

For comparison, we also integrated the full equation of motion (Eq.~\ref{eq:eomcart}) in Cartesian coordinates for the same AGN model and stochastic prescription. In this case, we adopted a second-order modified midpoint leapfrog integrator with a fixed timestep $\delta t = 10^{-4}P$, where $P$ is the orbital period of $m_{\rm BH}$ around $M_{\rm BH}$ \citep[see, e.g.,][]{mikkola2006}. After recovering the Keplerian elements from the Cartesian trajectories using standard formulae \citep[see, e.g.,][]{2005ormo.book.....R}, we compared the statistical outcomes of both approaches. For an ensemble of $2 \times 10^3$ orbits evolved over 20 unperturbed periods from identical initial conditions, we find that the spread in eccentricity and semimajor axis differs by approximately 5\% and 25\%, respectively. This discrepancy is attributed to the perturbative nature of the Keplerian-element equations. In contrast, the average values of $a$, $e$, and $\iota$ remain in good agreement between the two methods, with deviations consistently below 1\%.

\section{Simulations and results}\label{results}
\subsection{Evolution of the orbital elements}
Figure~\ref{fig:evolturb} shows the evolution of the orbital elements of a $20\,\msun$ BH embedded in the AGN disk, comparing cases with and without turbulent forcing. The BH is initially placed at the migration trap radius, $R \sim 10^4 \,\Rg \sim 10^3 \,\au$, where the net migration torque vanishes. It begins with an eccentricity of $e = 0.1$ and an inclination of $\iota = 0.5^\circ$,  \REV{ensuring that the orbit remains within a few disk scale heights. At this radius the disk has an aspect ratio of $H/R = 8.5 \times 10^{-3}$, corresponding to an inclination of $0.487^\circ$}. The BH has an initial period of $P = 9.1 \,\rm yr$.

In the absence of turbulence (blue curve), the BH rapidly circularizes and aligns with the disk midplane. During the first two orbital periods, dynamical friction is enhanced near the midplane crossings (vertical lines) due to the steep vertical density gradient. At each crossing, the semimajor axis increases when the BH passes through the disk from above and decreases when crossing from below. This asymmetry arises because the first crossing occurs near apocenter (corresponding to the negative $\cos{u}$ in Eq.~\ref{eq:linear}) where the BH's orbital velocity is lower than the local circular velocity, allowing it to gain energy from the gas flow. After approximately three crossings (i.e., ${\sim} 1.5$ orbital periods), the inclination drops below ${\sim} 0.18^\circ$ ($z_{\rm max}/H \sim 0.4$), such that the vertical variations in gas density become negligible. From this point on, the semimajor axis stabilizes at the final value of $a_{\rm fin} \simeq 975 \,\mathrm{au} \simeq 9.9 \times 10^4 \, \Rg$, while both eccentricity and inclination decay exponentially, consistent with the analytic expectations from Eqs.~(\ref{eq:epert}--\ref{eq:iotapert}) and the results of hydrodynamical simulations \citep{rowan2025b,whitehead2025b}.

Notably, the inclination decays at an exponential rate of ${\sim} \eta_*/2.6$, which is slightly slower than the rate ${\sim} \eta_*/2$ obtained by averaging the factor $\cos^2{u}$ in Eq.~(\ref{eq:iotapert}) over the argument of latitude $u$. This discrepancy likely arises because the inclination decays too rapidly for $u$ to circulate during the alignment phase. Given the eccentricity damping timescale $\tau_* = 1/\eta_* \simeq 0.75 \,\mathrm{yr} \simeq 0.08 P$, the orbit settles into the disk while $u$ librates around $u_*=5.38$ such that $1/\cos^2{u_*}\approx 2.6$, making the decay rate sensitive to the librating argument of latitude rather than its time-averaged behavior. 

In the joint limit of low eccentricity and inclination, the fixed point $u_*$ can be estimated by setting ${\rm d}u/{\rm d}t=0$, yielding
\begin{equation}
        n \left[1+2e\cos{(u-\omega_0)}\right] = \frac{1}{4} \eta_* \left[\cos(u-\omega_0) - 2\right] \sin{(2u)},
\end{equation}
where $n=\sqrt{{\mu}/{a^3}}$ is the mean motion and $\omega_0$ is the argument of pericenter at the onset of the exponential decay.
This transcendental equation admits real roots only in the fast damping regime $\eta_* \gg n$, i.e., when the damping timescale is shorter than the orbital period. As $n/\eta_* \rightarrow 0$, the solutions collapse onto the four zeros of $\sin(2u)=0$. However, only the zeros at $0$ and $\pi$ are physical because the inclination damping in Eq.~(\ref{eq:iotapert}) would otherwise vanish. Setting $e=0$, a first‐order expansion in $n/\eta_*$ about each root then gives
\begin{equation}\label{eq:ustar}
        u_*^{(k)} = k \pi + \frac{n}{\eta_*} \frac{2}{\cos(k\pi - \omega_0) -2},\quad k\in\{0,1\}.
\end{equation}
We obtain $u_*^{(3)} = 5.48$, in good agreement with the numerically determined libration of the argument of latitude. This level of agreement is notable given the relatively large value of the ratio $n/\eta_* \simeq 0.48$, which pushes the limits of the fast damping approximation used in the derivation. 

In the turbulent case (red curves), the evolution of the BH becomes stochastic once it settles into the disk at ${\sim} 1.5$ orbital periods. The semimajor axis begins to diffuse around its final unperturbed value $a_{\rm fin}$, as predicted by Eq.~(\ref{eq:linearturb}). Both eccentricity and inclination deviate from the deterministic case, exhibiting sustained fluctuations above zero. Over time, they reach a steady state in which stochastic forcing balances dynamical damping.
When ensemble-averaged (dotted black line), $e$ and $\iota$ converge to equilibrium values of $(e, \iota)_{\rm eq} \simeq (1.0\times10^{-3}, 1.3\times10^{-3})$. This behavior supports the statistical description we introduced in Sect.~\ref{sec:stochastic}, where the interplay between linear damping and stochastic driving yields steady-state Rayleigh distributions for both $e$ and $\iota$.

The dash-dotted green lines in the lower panels represent the order-of-magnitude theoretical equilibrium root mean square values of $e$ and $\iota$ based on the ratio between the turbulent velocity dispersion and the local gas circular velocity, $\sigma_{\rm turb}/v_{\rm circ} \simeq 8.6\times10^{-4}$ (see Eq.~\ref{eq:sigmaturb}). The good agreement with the mean values supports the idea that the final state reflects a statistical balance between stochastic driving and dissipative damping.

\subsection{Steady-state distributions of $e$ and $\iota$}

The turbulent velocity field hampers the full orbital circularization and alignment of the embedded objects. This occurs because dynamical friction tends to bring the BH at rest with respect to the local velocity field, which is dominated by large-scale turbulent flows. Consequently, the BH's orbit does not become fully circular or aligned with the midplane but retains a residual eccentricity and inclination. On average, the eccentricity and inclination settle onto a plateau, whose characteristic magnitude is determined by the ratio of the turbulent velocity dispersion $\sigma_\mathrm{t}$ and the local circular velocity $v_\mathrm{circ}$. This is shown in Fig.~\ref{fig:dist}, which depicts the distribution of the eccentricity and inclination of the embedded BHs after 20 orbital periods, roughly corresponding to $244 \,\tau_*$.\\
\indent The semimajor axis undergoes a random walk due to the stochastic forcing, whose mean square amplitude can be estimated from statistical considerations. Given the stochastic derivative in Eq.~(\ref{eq:linearturb}), the mean squared change $\langle \Delta a^2 \rangle$ is
\begin{align}\label{eq:deltaderiv}
        \left\langle \frac{{\rm d}a}{{\rm d}t}, \frac{{\rm d}a}{{\rm d}t} \right\rangle & = \left( \frac{2 a\eta_*}{v_\mathrm{circ}} \right)^2 \int_0^t\int_0^t v_\mathrm{turb}(t_1) v_\mathrm{turb}(t_2) {\rm d}t_1 {\rm d}t_2  ,\\
        & = \left( \frac{2 a\eta_* \sigma_{\rm turb}}{v_\mathrm{circ}} \right)^2 \int_0^t\int_0^t \exp{\left(-\frac{|t_1 - t_2|}{\tau_{\rm c}}\right)}{\rm d}t_1 {\rm d}t_2,
\end{align}
which for $t \gg \tau_{\rm c}$ reduces to
\begin{equation}\label{eq:deltaa}
        \langle \Delta a^2 \rangle(t) = 8 \eta_*^2 a^2 \frac{\sigma_\mathrm{turb}^2}{v_{\rm circ}^2}\, t,
\end{equation}
where we used $v^2_{\rm circ}=\mu/a$ and $\tau^2_{\rm c} = a^3/\mu$. We plot Eq.~(\ref{eq:deltaa}) as a function of time in Fig.~\ref{fig:evolturb} and show the corresponding normal distribution in Fig.~\ref{fig:dist}. Our analytic estimate matches the numerical results very well, despite the many assumptions, with small deviations likely due to neglecting the deterministic derivative in Eq.~(\ref{eq:fullpert}). In fact, even though any pure deterministic term would average to zero in Eq.~(\ref{eq:deltaderiv}), Eq.~(\ref{eq:fullpert}) depends on the eccentricity, whose mean square value is nonzero, albeit small.\\
\indent Contrary to what happens for the semimajor axis, the distributions of $e$ and $\iota$ settle to a steady state, reasonably well described in terms of Rayleigh distributions. In the middle and bottom panel of Fig.~\ref{fig:dist} we compare the distributions obtained from the numerical integrations (red histograms) with the matching Rayleigh distribution (solid green lines) with a choice of scale parameter so that the mean value is equal to $\sigma_{\rm turb}/v_{\rm circ}$. For both quantities, the empirical distribution and its semi-analytical estimate differ by less than the 15\% over the significant range of $e$ and $\iota$. We note that, for the distribution of eccentricity, an even better match is obtained by setting the mean value to  $1.2\sigma_{\rm turb}/v_{\rm circ}$.

\subsection{Stochastic equilibrium of $e$ and $\iota$}\label{sec:stochastic}

The equations for ${\rm d}e/{\rm d}t$ and ${\rm d}\iota/{\rm d}t$ exhibit similar dynamics. The deterministic terms in Eqs.~(\ref{eq:epert}--\ref{eq:iotapert}) include linear damping components proportional to $-\eta_* e$ and $-\eta_* \iota$, while the stochastic terms in Eqs.~(\ref{eq:eturb}--\ref{eq:iotaturb}) account for turbulent forcing, proportional to $v_{\rm turb}$. The deterministic terms drive $e$ and $\iota$ toward zero, while the stochastic terms introduce diffusion. We stress the fact that the above equations are valid in the subsonic regime where the friction coefficient reduces to $\eta_*$ (see Eq.~\ref{eq:etastar}). Over long timescales, the probability distributions for the eccentricity, $\mathcal{P}(e)$, and inclination, $\mathcal{P}(\iota)$, reach steady-state distributions dictated by the balance between frictional damping and stochastic forcing. These steady-state distributions can be derived analytically by exploiting the fact that Eqs.~(\ref{eq:eturb}--\ref{eq:lastlinearturb}) and (\ref{eq:epert}--\ref{eq:lastlinear}) are closely related to the general stochastic differential equation for a Ornstein–Uhlenbeck process $X_t$ (hereafter OU; see, e.g., \citealt{1989fpem.book.....R,1994hsmp.book.....G}) that reads
\begin{equation}\label{eq:ou}
        {\rm d}X_t = - \theta X_t {\rm d}t + \sigma {\rm d}W_t.
\end{equation}
In the equation above, $\theta$ is the damping rate, $\sigma$ is the noise intensity, and $W_t$ represents the underlying Wiener process. In our case $\eta_*$ and $\sigma_{\rm turb}$ correspond to $\theta$ and $\sigma$, respectively. At variance with the standard OU process, here we consider a Markovian stochastic forcing with an exponentially decaying autocorrelation function rather than a delta function.

We can explicitly recast the eccentricity evolution equations into the form of an OU process. While the equation for ${\rm d}e/{\rm d}t$ might appear OU-like, it is coupled to the evolution of the longitude of pericenter ${\rm d}\varpi/{\rm d}t$. To isolate the true linear stochastic equation, we switch variables to the components of the eccentricity vector:
\begin{align}
        &g_x = e \cos{\varpi},\\
        &g_y = e \sin{\varpi}.
\end{align}
After algebraic manipulation, we obtain
\begin{align}\label{eq:dgxdt}
        & \frac{{\rm d}g_x}{{\rm d}t} = - \eta_* g_x+ \eta_* \sqrt{\frac{a}{\mu}} \left( 2v_{{\rm turb}, \phi} \cos\ell + v_{{\rm turb}, r} \sin\ell \right), \\
        & \frac{{\rm d}g_y}{{\rm d}t} = - \eta_* g_y + \eta_* \sqrt{\frac{a}{\mu}} \left( 2v_{{\rm turb}, \phi} \sin\ell - v_{{\rm turb}, r} \cos\ell \right),
\end{align}
which describe a 2D OU-like process for the eccentricity vector. Here $\ell=\varpi + \nu = \Omega + \omega + \nu$ represents the true longitude. In the two equations above, the first term represents the deterministic linear damping $\theta$, and the second term is the stochastic forcing $\sigma$ due to the radial and azimuthal components of the velocity.

\begin{figure}
        \centering
        \includegraphics[width=\columnwidth]{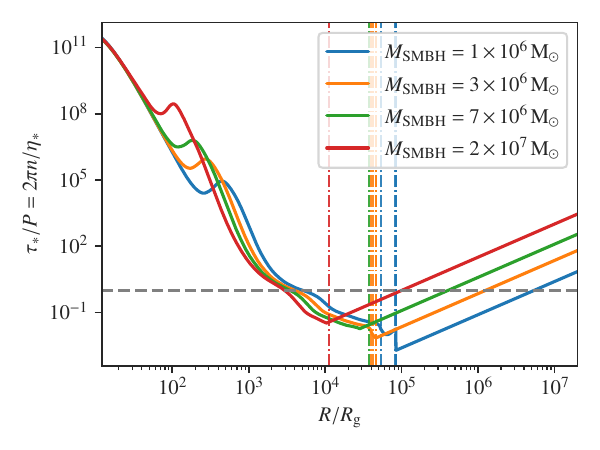}
        \caption{Ratio between the damping timescale $\tau_*$ (Eq.~\ref{eq:taustar}) and the orbital period $P$, as a function of the distance from the central SMBH. The estimate assumes a BH with mass  $m_{\rm BH}=20 \, \rm M_{\rm \odot}$ based on the AGN disk profile for SMBHs with masses $M_{\rm SMBH}=10^6, 3\times10^7, 7\times10^6,$ and $2\times 10^7 \, \rm M_{\rm \odot}$, with $\alpha = 0.1$. The vertical lines indicate the location of the migration traps.
        }
        \label{fig:taustar}
\end{figure}

A 1D OU process has a steady-state distribution that is a Gaussian with zero mean. 
The variance of this distribution can be derived using It\^o's calculus (e.g., see \citealt{1989fpem.book.....R,1994hsmp.book.....G}). We first expressed the solution of Eq.~(\ref{eq:dgxdt}) as It\^o's stochastic integral:
\begin{equation}
    g_x(t) = D \int^{t}_{-\infty} \exp{\Big(-\eta_* (t-s)\Big)} \,\Sigma_x(s) \,{\rm d}s,
\end{equation}
where $D = \eta_* / v_{\rm circ}$ and $\Sigma_x(t)=2v_{{\rm turb}, \phi} \cos{u(t)} + v_{{\rm turb}, r} \sin{u(t)}$. The variance of $g_x$ is then
\begin{align}
        \langle g_x^2 \rangle 
        &= D^2 \int_0^\infty\int_0^\infty \exp\Big( -\eta_*(t_1 + t_2) \Big) \cdot \notag \\
        &\qquad\qquad\qquad \left\langle \Sigma_x(t - t_1), \Sigma_x(t - t_2) \right\rangle \, {\rm d}t_1 \, {\rm d}t_2 = \notag \\
        &= \frac{2 \eta_* \sigma^2_{\rm turb}}{v^2_{\rm circ}} \int_0^\infty \exp(-\eta_* t) S(t) \, {\rm d}t = \notag\\
        & = \frac{2 \eta_* \sigma^2_{\rm turb}}{v^2_{\rm circ}}  \frac{A \,(\eta_* + \Omega_{\rm circ}) - B \,n}{(\eta_* + \Omega_{\rm circ})^2 + n^2},
 \end{align}
where
\begin{align}
& S(t) =  \exp{(-t/\tau_c)} \left[A \cos(nt) - B \sin(nt) \right],\\
& A =  4\cos^2\ell + \sin^2\ell, \\ 
& B =  3\cos\ell\sin\ell.
\end{align}
We distinguish $\tau_{\rm c} = 1 / \Omega_{\rm circ}$, the correlation time of turbulent forcing (set by the gas), from the BH's mean motion $n$. Although they are equal in our model, this separation allows us to keep the generality for more complex models. 

The $A$ and $B$ terms depend on the true longitude and need to be estimated depending on the ratio between the damping rate $\eta_*$ and the orbital and gas correlation frequencies $n$ and $\Omega_{\rm circ}$. Figure~\ref{fig:taustar} shows how the damping timescale $\tau_*$ compares to the orbital period in the AGN disk.

In the region of the migration trap at $R\sim 10^4 \, R_{\rm g}$, we obtain $n/\eta_* \simeq 0.48$. In this case, the damping acts moderately faster than the orbital timescale, and we consider the true longitude $\ell$ to be constant over one damping timescale. However, since $\ell$ continues to circulate and is not dynamically driven to a fixed point (unlike the argument of latitude $u$, which librates around $u_*$, see Eq.~\ref{eq:ustar}), we do not expect any preferred value of $\ell$ during the damping. We can thus ensemble average $A$ over $\ell$, yielding $\langle A \rangle = 5/2$. Using $\sigma^2_e = 2 \langle g^2_x\rangle$. This holds due to the statistical symmetry between $g_x$ and $g_y$. The steady-state variance of the eccentricity in the fast damping regime reads:
\begin{equation}\label{eq:sigmae_fast}
\sigma^2_e \simeq  10 \left(\frac{\sigma_{\rm turb}}{v_{\rm circ}}\right)^2 \quad \text{(fast damping)}.
\end{equation}

In other regions of the AGN disk, or for smaller BHs, the system can enter the slow damping regime where $\eta_* \ll n$ . In this case, the true longitude circulates rapidly compared to the damping timescale, and we can average over its values. As a result, the stochastic forcing averages over all phases, leading to the orbit-averaged values of $\langle A \rangle = 5/2$ and $\langle B \rangle = 0$. The resulting eccentricity variance in the slow damping regime is
\begin{equation}\label{eq:sigmae_slow}
        \sigma^2_e \simeq 5\,\frac{\eta_*}{n} \left(\frac{\sigma_{\rm turb}}{v_{\rm circ}}\right)^2 \quad \text{(slow damping)},
\end{equation}
where we set $\Omega_{\rm circ} = n$. This expression shows that the stochastic excitation is suppressed by the small parameter $\eta_* / n \ll 1$, reflecting the fact that the orbital motion rapidly de-correlates each turbulent kick before it can accumulate significant eccentricity.

The variance of the inclination, $\sigma_\iota^2$, can be derived by applying the same OU process analysis to the inclination vector components obtained via the change of variables:
\begin{align}
    &h_x = i \cos{\Omega},\\
    &h_y = i \sin{\Omega}.
\end{align}
In this formulation, $h_x$ and $h_y$ follow stochastic differential equations analogous to those for eccentricity. The resulting steady-state variance of $\iota$ takes the form:
\begin{equation}
\sigma^2_\iota =
        \begin{cases}
                \dfrac{2}{3} \left(\dfrac{\sigma_{\rm turb}}{v_{\rm circ}}\right)^2, & \text{if } \eta_* \gg n \quad \text{(fast damping)},\\[2.5ex]
                2 \dfrac{\eta_*}{n} \left(\dfrac{\sigma_{\rm turb}}{v_{\rm circ}}\right)^2, & \text{if } \eta_* \ll n \quad \text{(slow damping)}.
        \end{cases}
\end{equation}
These analytical results are consistent with the numerical simulations shown in Fig.~\ref{fig:dist}, and they account for the observed difference between $\sigma_\iota$ and $\sigma_e$. In particular, they explain why the steady-state variance of eccentricity exceeds that of inclination. For the simulations considered here, the ratio $n/\eta_* \simeq 0.48$ places the system between the fast and slow damping regimes, leading to intermediate values of $\sigma_\iota$ and $\sigma_e$.

\section{Summary and conclusions}\label{discussion}
Using semi-analytical stochastic methods and numerical simulations, we investigated the drag exerted by a turbulent AGN disk surrounding an SMBH on a stellar-mass BH initially located at the migration trap. We introduced a numerical technique, inspired by analogous studies in protoplanetary disks, that incorporates turbulence, modeled as a stochastic velocity field, without solving the full hydrodynamical equations. Instead, the method relies on effective stochastic differential equations.

We implemented a full solver in Cartesian coordinates and a reduced solver based on Keplerian orbital elements, and find excellent agreement between the two approaches across a range of initial conditions. In the subsonic regime ($\mathcal{M} \ll 1$), we derived approximate Langevin equations governing the orbital element evolution, enabling us to compute the stationary distributions of eccentricity $e$ and inclination $\iota$. We compared the resulting distributions from $N = 2.5 \times 10^4$ independent realizations to analytical Rayleigh probability density functions, finding good agreement.

We analytically estimated the variance of the steady-state Rayleigh distributions using the linearized stochastic equations and find good agreement with the results of the full numerical simulations. By expressing the relevant quantities in terms of local disk properties, we obtain the following expressions:
\begin{equation}\label{eq:finalsigma}
\boxed{
\begin{aligned}
        & \sigma_e^2 = 15\, \sigma_\iota^2 = 10\, \alpha \left(\frac{H}{R}\right)^2,
        \quad \text{(fast damping, } \eta_* \gg n),\\
        & \sigma_e^2 = \frac{5}{2}\, \sigma_\iota^2 = 5\, \alpha \frac{\eta_*}{n} \left(\frac{H}{R}\right)^2, 
        \quad \text{(slow damping, } \eta_* \ll n).
\end{aligned}
}
\end{equation}
Here, $\alpha$ is the alpha-viscosity parameter characterizing the strength of turbulence, $H$ is the disk scale height, $n = \Omega_{\rm circ} = \sqrt{G M_{\rm SMBH}/R^3}$ is the circular orbital frequency, and $\eta_*$ is the subsonic damping rate determined by the local gas properties (see Eq.~\ref{eq:etastar}).

Equation~(\ref{eq:finalsigma}) can be readily used to sample the eccentricity and inclination of BHs embedded in AGN disks, both in Monte-Carlo simulations \citep{mckernan2020b,tagawa2021,rowan2024a,mcfacts1,mcfacts2,mcfacts3} and as initial conditions for BH scattering hydrodynamical simulations \citep{whitehead2025,rowan2025}, which to date have assumed perfectly circular and aligned orbits. 
Notably, \citet{trani2024a} show that the initial eccentricity and inclination of BHs in a disk configuration strongly affect the merger rate from three-body encounters. Specifically, they find that the merger rate in a dispersion-dominated disk is suppressed by a factor of ${\sim}42$ relative to a shear-dominated disk (see their figure~5). A similar suppression is expected for the formation of binaries via Jacobi captures mediated by dynamical friction, since even modest values of $e$ and $\iota$ increase the relative velocity between interacting BHs, significantly raising the amount of energy that must be dissipated for capture to occur \citep[see][]{dodici2024}.

In our current formulation and numerical implementation, we neglected migration torques arising from the perturbation of the disk’s density by the orbiting BH. As a result, our simulations and theoretical predictions likely overestimate the diffusion in semimajor axis. In reality, such torques would act as a restoring force, counteracting orbital changes from the gas drag as well as the stochastic fluctuations in $a$. Consequently, the BH would remain near the migration trap, and the semimajor axis distribution would settle into a steady-state centered on the trap location. Nonetheless, our simulations show that the radial wandering remains sufficiently small to justify the assumption of constant $a$ used in the derivation of Eq.~\ref{eq:finalsigma}. 

We note that simplified hydrodynamical simulations by \citet{wu2024} suggest that turbulence may also reduce the magnitude of the migration torque itself, thereby diminishing its ability to stabilize the semimajor axis \citep[see also][]{baruteau2010,pierens2012,stoll2017}. We stress that the framework presented here relies on the linear theory of dynamical friction, which neglects nonlinear interactions between the BH’s wake and the turbulent gas. Furthermore, our treatment neglected the back-reaction of the embedded BH on the surrounding gas. In reality, accretion onto the BH can heat the local disk, drive local turbulence, or create a low-density cavity, potentially reducing the drag and altering the migration trap structure \citep[e.g.,][]{gilbaum2022,epsteinmartin2025,chen2025}. Quantifying these effects lie beyond the scope of this work and should be investigated using hydrodynamical simulations that self-consistently model both turbulence and gas--BH coupling.
\begin{acknowledgements}
A.A.T. would like to thank Martin Pessah and Troels Haugbølle for helpful discussions, which provided valuable directions for this research. A.A.T. further acknowledges the hospitality of Firenze University's physics \& astronomy department at Arcetri, where this work was started, and acknowledges support from the Horizon Europe research and innovation programs under the Marie Sk\l{}odowska-Curie grant agreement no. 101103134. P.F.D.C. wishes to acknowledge funding by ``Fondazione Cassa di Risparmio di Firenze'' under the project {\it HIPERCRHEL} for the use of high performance computing resources at the University of Firenze. We thank the referee for their insightful comments, which helped improve the clarity of this work.
\end{acknowledgements}
\bibliographystyle{aa} 
\bibliography{dragfric} 
\end{document}